# MODELING OF THERMAL MAGNETIC FLUCTUATIONS IN NANOPARTICLE ENHANCED MAGNETIC RESONANCE DETECTION

Tahmid Kaisar[1], Md Mahadi Rajib[1], Hatem ElBidweihy[2], Mladen Barbic[3] and Jayasimha Atulasimha[1*]

[1]*Department of Mechanical and Nuclear Engineering, Virginia Commonwealth University, Richmond, VA 23284, USA*
[2]*United States Naval Academy, Electrical and Computer Engineering Department, Annapolis, Md 21402, USA*
[3]*NYU Langone Health, Tech4Health Institute, New York, NY 10010, USA*
\* Corresponding author: jatulasimha@vcu.edu

**Abstract**

We present a systematic numerical modeling investigation of magnetization dynamics and thermal magnetic moment fluctuations of single magnetic domain nanoparticles in a configuration applicable to enhancing inductive magnetic resonance detection signal to noise ratio (SNR). Previous proposals for oriented anisotropic single magnetic domain nanoparticle amplification of magnetic flux in MRI coil focused only on the coil pick-up voltage signal enhancement. Here we extend the analysis to the numerical evaluation of the SNR by modeling the inherent thermal magnetic noise introduced into the detection coil by the insertion of such anisotropic nanoparticle-filled coil core. We utilize the Landau-Lifshitz-Gilbert equation under the Stoner-Wohlfarth single magnetic domain (macrospin) assumption to simulate the magnetization dynamics in such nanoparticles due to AC drive field as well as thermal noise. These simulations are used to evaluate the nanoparticle configurations and shape effects on enhancing SNR. Finally, we explore the effect of narrow band filtering of the broadband magnetic moment thermal fluctuation noise on the SNR. Our results provide the impetus for relatively simple modifications to existing MRI systems for achieving enhanced detection SNR in scanners with modest polarizing magnetic fields.

**Introduction**

Sensitivity enhancement in magnetic resonance detection continues to be an important challenge due to the importance of NMR and MRI in basic science, medical diagnostics, and materials characterization [1-5]. Although many alternative methods of magnetic resonance detection have been developed over the years, inductive coil detection of magnetic resonance of precessing proton nuclear magnetic moments is by far the most common [6]. The challenge in magnetic resonance detection stems from the low nuclear spin polarization at room temperature and laboratory static magnetic fields. An additional challenge is the fundamental requirement that the detector in magnetic resonance experiment needs to be compatible with and immune to the large polarizing DC magnetic field while also sufficiently sensitive to weak AC magnetic fields generated by the precessing nuclear spins. The inductive coil, operating on the principle of Faraday's law of induction, satisfies this requirement, and enhancing the inductive coil detection SNR has been pursued through various techniques [7-9]. However, unlimited increase of the polarizing magnetic field is cost prohibitive, and technical challenges often inhibit the development of mobile MRI units, their access, sustainability, and size. Therefore, solutions to achieving sufficient or improved SNR in NMR inductive coil detection in lower magnetic fields and more accessible and compact configurations remains highly desirable [10].

**Signal amplification by magnetic nanoparticle-filled coil core**

An idea has been put forward to increase the magnetic field flux from the sample through the coil by filling the coil with a core of oriented anisotropic single domain magnetic nanoparticles [11, 12], as shown in Figure 1. The sample and the inductive coil detector are both in the prototypical MRI environment of a large DC polarizing magnetic field along the z-axis, $B_{Zdc}=\mu_0 H_{Zdc}$, where $\mu_0$ is the permeability of free space. This field generates a fractional nuclear spin polarization of protons in the sample. Application of RF magnetic fields along the x-axis is subsequently used to tilt the magnetic moment of the sample away from the z-axis and generate precession of the sample magnetization around the z-axis at the proton NMR frequency $\omega_0=\gamma B_{Zdc}$, where $\gamma$ is the proton nuclear gyromagnetic ratio. This sample moment precession around the z-axis generates a time-varying magnetic field $B_{Xac}=\mu_0 H_{Xac}$ through the inductive coil detector of N turns and sensing area A along the x-axis. By Faraday's law of induction, an AC signal voltage V at frequency $\omega_0$ generated across the coil terminals is:

$$V = N \cdot A \cdot \omega_0 \cdot B_{Xac} \tag{1}$$

It is a well-known practice in electromagnet design and ambient inductive detectors that a soft ferromagnetic core within the coil significantly amplifies the magnetic flux through the coil [13, 14]. The challenge, however, in the configuration of NMR detection of Figure 1 is that the presence of the large polarizing magnetic field along the z-axis, $B_{Zdc}=\mu_0 H_{Zdc}$, would generally saturate the detection coil core made of a soft ferromagnet along the z-axis and render the AC magnetic field due to proton precession along the x-axis of the coil ineffective. The solution proposed [11, 12] was that the oriented anisotropic magnetic nanoparticles filling the coil core actually have an appreciable magnetic susceptibility along the x-axis precisely in the presence of a significant DC magnetic field along the z-axis. The pick-up coil voltage is then:

$$V = N \cdot A \cdot \omega_0 \cdot \mu_0 \cdot (H_{Xac} + M_{Xac}) \quad (2)$$

Where $M_{Xac}$ is the magnetization component of the nanoparticle-filled coil core along the x-direction (sensing direction of the coil) due to the magnetic field $B_{Xac}=\mu_0 H_{Xac}$ from the precessing sample nuclear spin moment, $M_{Xac}=\chi_{RT}H_{Xac}$ (where $\chi_{RT}=\Delta M_{Xac}/\Delta H_{Xac}$ is defined as reversible transverse susceptibility). Therefore, if the reversible transverse susceptibility, $\chi_{RT}$, of the magnetic nanoparticle-filled coil core along the x-axis is significant at the large polarizing DC magnetic field $B_{Zdc}$ along the z-axis, the inductive coil signal voltage will be enhanced. Various theoretical [15-17] and experimental investigations [18-25] of reversible transverse susceptibility in oriented magnetic nanostructures indeed reveal that its magnitude can be appreciable and therefore might provide a viable route for magnetic resonance signal amplification, as diagrammatically shown in Figure 1.

In this work, we numerically evaluate the coil signal voltage by modeling individual nanoparticle magnetic moment dynamics in the Stoner-Wohlfarth uniform magnetization approximation [26]. More specifically, we evaluate the AC nanoparticle moment along the x-axis in Figure 1, $m_{Xac}$, in the presence of a large DC magnetic field $B_{Zdc}=\mu_0 H_{Zdc}$ along the z-axis and under the driven sample AC magnetic field $B_{Xac}=\mu_0 H_{Xac}$ along the x-axis. We assume for simplicity that the total coil core of volume, $v_C$, is composed of "n" number of identical oriented single domain magnetic nanoparticles, and that each particle x-component of the AC magnetic moment, $m_{Xac}$, equally contributes to the coherent amplification of the pick-up voltage signal of the coil detector. Therefore, the total coil AC voltage due to the magnetic nanoparticle core contribution is:

$$V = N \cdot A \cdot \omega \cdot \mu_0 \cdot n \cdot \frac{m_{Xac}}{v_C} \quad (3)$$

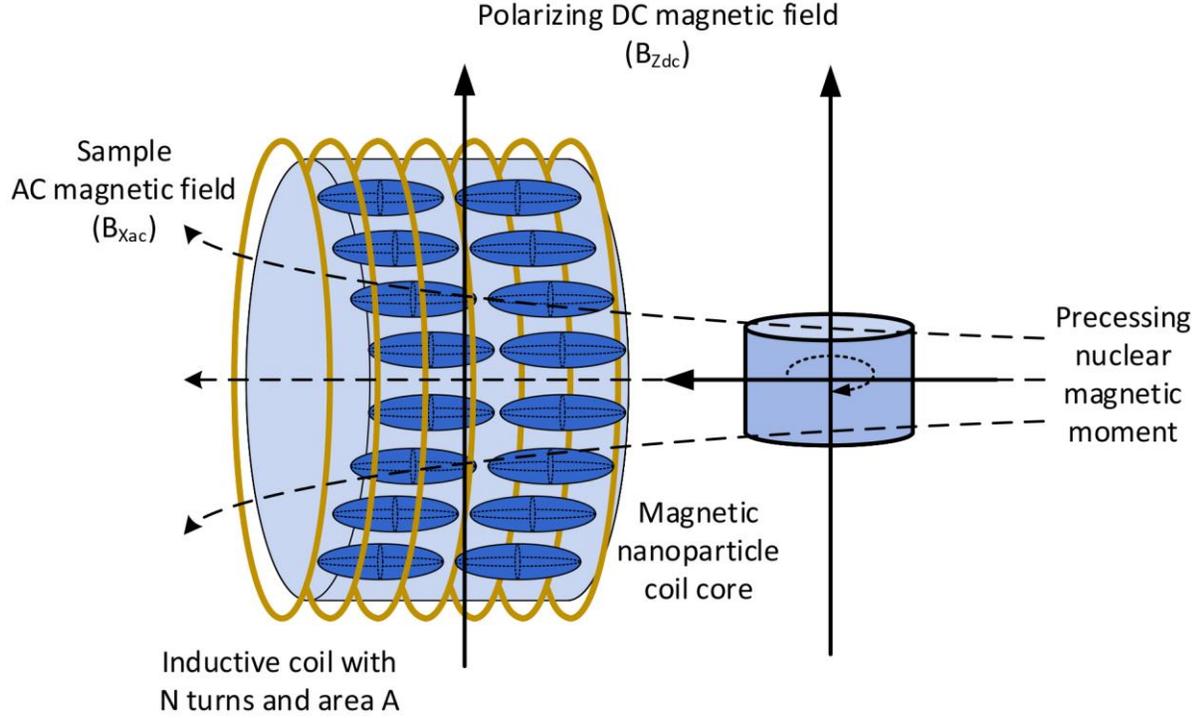

FIG. 1. Schematic diagram for NMR detection with magnetic nanoparticle filled coil core.

**Noise contribution by magnetic nanoparticle-filled coil core**

Essential to the SNR consideration of any NMR experimental arrangement is the evaluation of the noise sources in the signal chain. Since in this work we focus specifically on the magnetic nanoparticle-filled inductive coil detector, we will not consider the sample noise, the amplifier noise, and the Johnson noise contributions which are addressed in numerous works [27-31]. Any magnetic material placed inside the inductive detection coil will introduce additional pick-up voltage noise due to intrinsic magnetization fluctuations [32]. In our specific case, diagrammatically shown in Figure 1, these thermal fluctuations of the coil core magnetization along the x-axis, which we numerically model in detail in this work, generate a total mean squared coil noise voltage:

$$<V^2> = N^2 \cdot A^2 \cdot \omega^2 \cdot \mu_0^2 \cdot <M_X^2> \qquad (4)$$

We assume for simplicity that the total coil core of volume, $v_C$, is composed of n number of identical oriented single domain magnetic nanoparticles, and that each particle magnetic moment, m, undergoes random uncorrelated thermal fluctuation. Therefore, the total mean squared coil noise voltage due to the magnetic nanoparticle core is [33, 34]:

$$<V^2> = N^2 \cdot A^2 \cdot \omega^2 \cdot \mu_0^2 \cdot n \cdot \frac{<m_X^2>}{v_C^2} \tag{5}$$

We note that the magnetic moment fluctuation phenomena has previously been investigated in various spin systems, materials, and detection modalities [35-42]. However, we are not aware of any theoretical, numerical, or experimental study where thermal magnetic moment fluctuations of single domain nanoparticles of the configuration of Figure 1 (where the large polarizing magnetic field is applied perpendicular to the nanoparticle hard axis and the coil detection axis) and their contribution to the coil noise voltage has been carried out.

In this article, therefore, in order to assess the signal and the noise of the configuration of the magnetic resonance coil detector of Figure 1, we simulate the room temperature magnetization dynamics of a single domain nanomagnet in the coil core. Importantly, we also model the magnetic nanoparticle moment thermal fluctuation under the same conditions. We explore such single nanomagnet dynamics for the macrospin Stoner-Wolhfarth model uniform magnetization oblate and prolate ellipsoid geometries. This will explain the optimum nanoparticle orientation and bias field needed to maximize SNR of the experimental arrangement of Figure 1. We then extend this analysis to scaling properties of an ensemble of nanomagnets and study the effect of applying a band-pass filter to provide an estimate on the extent to which the insertion of magnetic particles in the sensing coil can enhance the limits of detection of magnetic fields due to proton spin resonances in MRI/NMR.

**Modelling particle magnetization dynamics in the presence of room temperature thermal noise**

Modelling of the single particle magnetization dynamics was performed by solving the Landau-Lifshitz-Gilbert (LLG) equation [43]:

$$\frac{\partial \vec{m}}{\delta t} = \left(\frac{-\gamma}{1+\alpha^2}\right)[\vec{m} \times H_{eff} + \alpha\{\vec{m} \times (\vec{m} \times \vec{H}_{eff})\}] \tag{6}$$

In equation (6), $\gamma$ is the gyromagnetic ratio (rad/Ts), $\alpha$ is the Gilbert damping coefficient and $\vec{m}$ is the normalized magnetization vector, found by normalizing the magnetization vector ($\vec{M}$) with respect to saturation magnetization ($M_s$). The effective field ($\vec{H}_{eff}$) was obtained from the derivative of the total energy (E) of the system with respect to the magnetization ($\vec{M}$):

$$\vec{H}_{eff} = -\frac{1}{\mu_0 \Omega}\frac{dE}{d\vec{M}} + \vec{H}_{thermal} \tag{7}$$

where $\mu_0$ is the permeability of the vacuum and $\Omega$ is the volume of the nanomagnet.

The total potential energy in equation (7) is given by:

$$E = E_{shape\ anisotropy} + E_{zeeman} \quad (8)$$

where $E_{shape\ anisotropy}$ is the shape anisotropy due to the prolate or oblate shape and can be calculated from the following equation:

$$E_{shape\ anisotropy} = \left(\frac{\mu_0}{2}\right)\Omega[N_{d\_xx}M_x^2 + N_{d\_yy}M_y^2 + N_{d\_zz}M_z^2] \quad (9)$$

with $N_{d\_xx}$, $N_{d\_yy}$, and $N_{d\_zz}$ representing the demagnetization factors in the respective directions of the nanomagnet and $E_{zeeman}$ is the potential energy of nanomagnet for an external magnetic field ($\vec{H}$), given by:

$$E_{zeeman} = -\mu_0 \Omega \vec{H} \cdot \vec{M} \quad (10)$$

The thermal field $\vec{H}_{thermal}$ is modeled as a random field incorporated in the manner of [44, 45]

$$\vec{H}_{thermal}(t) = \sqrt{\frac{2kT\alpha}{\mu_0 M_s \gamma \Omega \Delta t}}\ \vec{G}(t) \quad (11)$$

In equation (11), $\vec{G}(t)$ is a Gaussian distribution with zero mean and unit variance in each Cartesian coordinate axis, k is the Boltzman constant, T is the temperature, $M_s$ is saturation magnetization and $\gamma$ is Gyromagnetic ratio. $\Delta t$ is the time-step used in the numerical solution of equation (6) and was chosen to be 100 fs. This was chosen to be small enough to ensure that all results are independent of the time step. We employed Euler method to solve the differential equations [44, 45].

Table I lists the values of the material properties of the nanomagnet.

**Table 1:** Material properties of CoFe [46]

| Parameters | Material Property |
|---|---|
| Saturation Magnetization ($M_s$) | **1.6x $10^6$ (A/m)** |
| Gilbert Damping (α) | **0.05** |
| Gyromagnetic Ratio (γ) | **2.2x10$^5$(m/As)** |

**Results and discussions**

Consider a prolate ellipsoid of volume ~5000 nm$^3$ (100nm×10nm×10nm) shown in Figure 2(a) where we assume the sample proton spin precession produces a magnetic field along the easy (long) x-axis of the nanomagnet while DC bias field is applied along one of the hard (short) axes, viz. the z-axis. When the DC bias field is zero there are two deep energy wells at θ=0, 180˚. When the magnetization is in one of these states, the magnetization response to an AC magnetic field along the x-axis (a simplified representation of the signal at the pick-up coil due to proton spin precession of Figure 1) is very small as the magnetization is in this deep potential as seen in Figures 3(a)-(b) and Table II. The corresponding magnetization fluctuation due to room temperature thermal noise (that is modelled as a random effective magnetic field, see equation 7) is also very small.

As the DC field increases along the z-axis to the point $H_{dc}=H_c$ (the DC bias field is equal to the coercive field $H_c$) the mean magnetization orientation is at 90˚. However, the potential well at 90˚ (Figure 2(a)) is characterized by a flat energy profile where the energy is nearly independent of the polar angle θ, around θ=90˚. This leads to a large magnetization response along the x-axis to an applied AC magnetic field along the x-axis (Figures 3(a)-(b) and Table II) in the presence of a large DC magnetic field along the z-axis. Importantly, since the energy profile is flat in this configuration, the particle moment fluctuation due to room temperature thermal noise is also high. Nevertheless, we find the Signal to Noise Ratio (SNR) is highest at $H_{dc}=H_c$. In fact, the magnetization response and the SNR ratio are found to increase monotonically with the applied bias field up to $H_c$ (see Table II based on the selected simulations shown in Figures 3(a)-(b) and all the simulations shown in Supplementary Figure S1) and then decreases as $H_{dc}>H_c$ (for example at $H_{dc}=1.25H_c$ in Table II and Supplementary Figure S1) due to an energy well deepening at θ=90˚ for $H_{dc}>H_c$ (Figure 2(a)). We note that the SNR is calculated in the following manner. First, we do not include any thermal noise and perform the LLG simulation to determine the magnetization response due to only the AC field from sample proton spin precession. Then we perform another LLG simulation with no signal and study the magnetization fluctuation due to the thermal noise only. The ratio of the rms values of the response due to the sample precessing field and that due to noise is defined as the SNR.

We note that the AC drive magnetic field along the z-axis has an amplitude of 800 A/m (10 Oe or 1 mT) for all cases discussed in this work, which is much larger than the typical signal due to proton spins which may be several orders of magnitude smaller. However, we chose a higher drive amplitude as we wanted to elicit a reasonable magnetization response that would be easily visible in the plots and result in reasonable SNR ratios for a single nanomagnet. In practise, the number of nanomagnets placed in the detector coil could be over n~$10^{12}$ or more resulting in sub nT sensing capability as discussed later. Furthermore, the proton resonance of 42.5 MHz occurs at a DC field ~ 1T, so the frequency of spin precession would be different at different anisotropy fields as the Zeeman (DC bias field) seen by the nanoparticles is the same as that seen by the proton spins in the MRI sample. However, to keep the simulations consistent, we assume signal due to proton resonance as 42.5 MHz for all cases as this would not change our qualitative findings.

Next, we consider a prolate ellipsoid shown in Figure 2(b) where we assume the proton spin precession produces a magnetic field along one of the hard (short) axes, viz. the x-axis of the nanomagnet while a DC bias field is applied along the long (easy) z-axis. Initially, the magnetization points downward ($\theta=180°$) and at $H_{dc}=0$ is in a deep potential well (Figure 2(b)). As the magnetic field applied along the +z direction ($\theta=0°$) increases, the energy well around $\theta=180°$ is flattened. Thus, for higher field, the magnetization response increases, as does the magnetization fluctuation due to thermal noise as shown in Table II (detailed simulations are all shown in the supplementary Figure S2). However, the shallow wells improve the magnetization response to the drive magnetic field more than the increased magnetization fluctuations due to thermal noise (as in the prior case) and increase the overall SNR ratio (Table II).

However, as one approaches $H_{dc}=H_c$ the SNR drops significantly. This is because the energy profile is flat at $\theta=180°$ but even small perturbations from this angle make the magnetization switch and rotate to the +z-axis ($\theta=0°$). Once it reaches this state, the energy well profile at $H_{dc}=H_c$ at $\theta=0°$ is a deeper than the well at $H_{dc}=0$. The reason is that the Zeeman energy due to field along +z-axis makes the already deep shape anisotropy well even deeper at $\theta=0°$ reducing both the magnetization response and the magnetization fluctuations due to thermal noise, as well as the SNR ratio. Thus, the best SNR is seen at $H_{dc}<H_c$ (see the high SNR at $0.875H_c$ in Table II) but close to $H_c$.

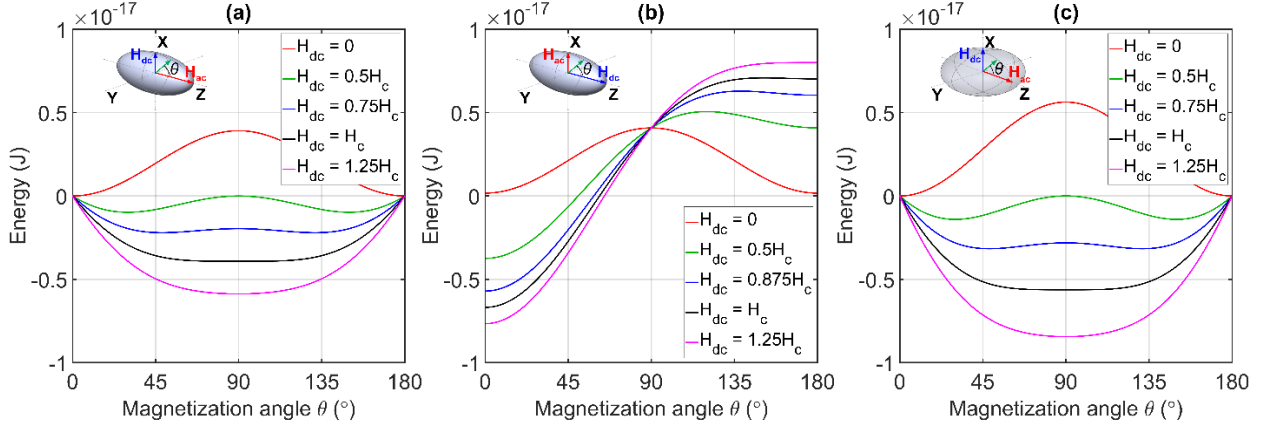

FIG. 2. Energy profile for various DC bias magnetic field ($H_{dc}$) for (a) Prolate: Bias field along minor axis, (b) Prolate: Bias field along major axis and (c) Oblate: Bias field along minor axis.

Finally, we consider the case of an oblate ellipsoid of volume ~5000 nm$^3$ (40nm×40nm×6nm), as same as the volume of prolate ellipsoid, shown in Figure 2(c) where we assume the proton spin precession produces a magnetic field along one of the easy (long) x-axis of the nanomagnet while DC bias field is applied along the hard z-axis. The symmetry of the problem is such that at $H_{dc}$=0 the magnetization is free to rotate in the x-y plane as there is no energy barrier to such rotation. As $H_{dc}$ increases, the magnetization is still free to move in a cone of the x-y plane at a specific angle to the z-axis that decreases with increasing $H_{dc}$, finally coinciding with it when $H_{dc}=H_c$. Thus, at a range of DC bias fields (for example from $H_{dc}$=0.25$H_c$ to $H_{dc}$=0.75 $H_c$) we observe a high SNR >1.4 when a single nanomagnet is driven by an AC magnetic field. This is due to the combination of high magnetization response given the symmetry and noise limited by presence of the DC bias field. The simulations of magnetic response to the AC magnetic field and magnetization fluctuations due to random thermal noise are respectively shown in Figures 3(c)-(d) comparing the $H_{dc}$=0 and $H_{dc}$=0.625$H_c$ cases with all other bias field cases shown in Figure S3 of the supplement.

In summary, as far as the SNR is concerned, the prolate ellipsoid with DC bias magnetic field along the hard axis (Figure 1(a)) is the better choice over the prolate ellipsoid with DC bias magnetic field along the easy axis. However, the oblate geometry and configuration shown in Figure 2(c) produces the highest SNR (more than twice higher than the highest SNR for the prolate ellipsoid configuration in Figure 2(a), and more than 10 times for the prolate ellipsoid configuration of Figure 2(b)). What makes this oblate configuration even more attractive to detection coils in MRI/NMR applications is that the high SNR performance is seen over a large range of DC bias field (e.g. $H_{dc}$=0.25$H_c$ to $H_{dc}$=0.75 $H_c$), making it attractive for broad range of MRI scanner fields.

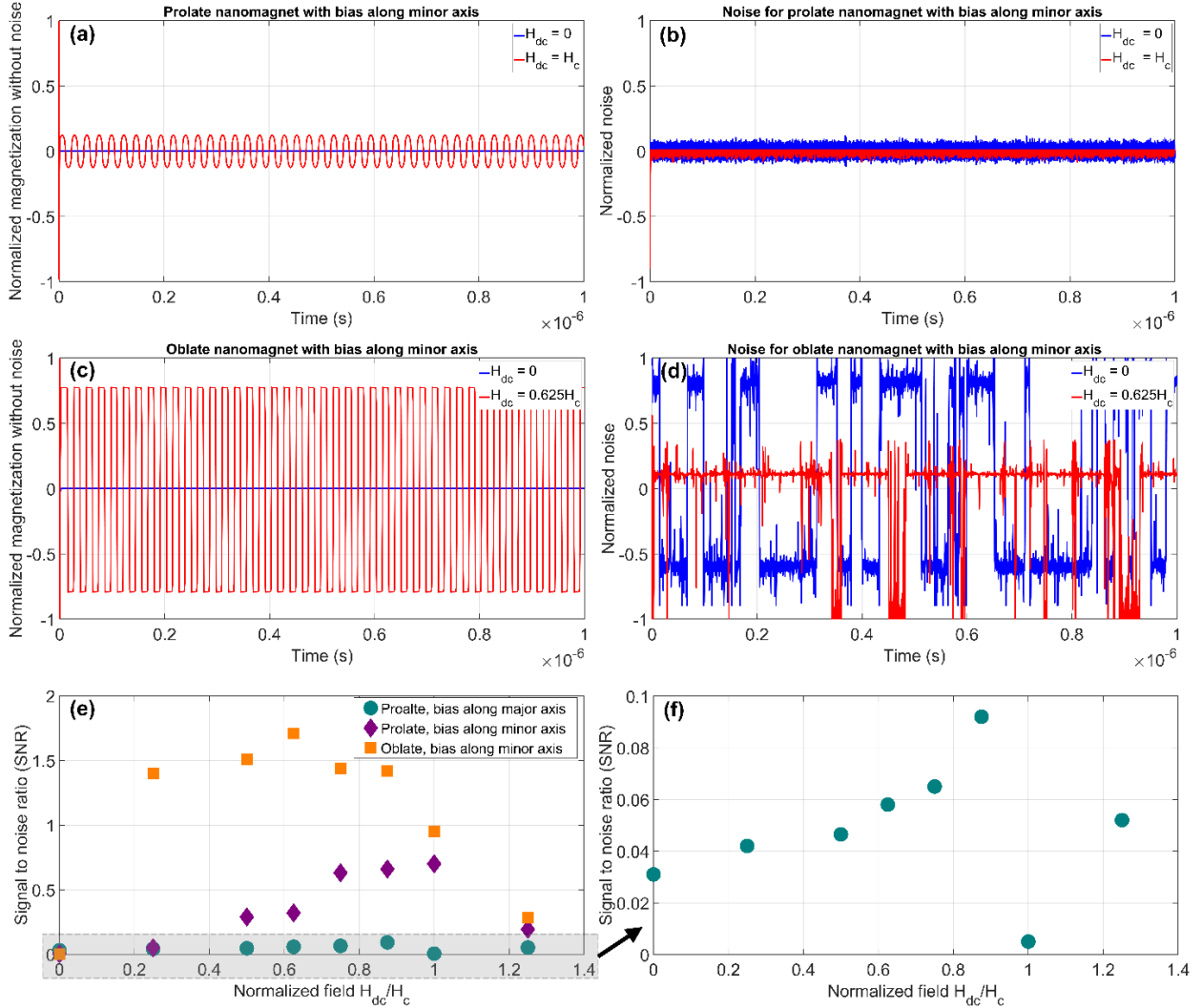

FIG. 3. Magnetization dynamics with (a) 800 A/m, 42.5 MHz field with no thermal noise for prolate nanomagnet with bias along minor axis and (b) only thermal noise at $H_{dc}=0$ and $H_{dc}=H_c$. Magnetization dynamics with (c) 800 A/m, 42.5 MHz field with no thermal noise for oblate nanomagnet with bias along minor axis and (d) only thermal noise at $H_{dc}=0$ and $H_{dc}= 0.625H_c$. (e) SNR vs $H_{dc}/H_c$ for prolate and oblate cases and (f) zoomed version of SNR vs $H_{dc}/H_c$ for prolate nanomagnet with bias along major axis.

**Table II:** Magnetization oscillations in the single nanomagnet of different geometries for different values of DC bias magnetic field (based on simulations shown in Figure 3 of the main paper and Figures S1, S2, S3 of the supplement).

| Cases | Value of bias field ($H_{dc}$) | RMS normalized magnetization ($M/M_s$) for a sinusoidal magnetic signal of 800 A/m (10 Oe) amplitude | RMS normalized magnetization ($M/M_s$) due to thermal noise only (no signal) | SNR (defined here as ratio of Column 2 and Column 3). |
|---|---|---|---|---|
| Prolate applying bias field along minor axis | 0 | $1.66 \times 10^{-6}$ | $6.45 \times 10^{-4}$ | .003 |
| | 0.25 $H_c$ | $2.82 \times 10^{-4}$ | .0075 | .05 |
| | 0.5 $H_c$ | 0.002 | 0.0103 | 0.2885 |
| | 0.75 $H_c$ | .0128 | 0.0202 | 0.63 |
| | $H_c$ | **0.1035** | **0.1464** | **.7042** |
| | 1.25 $H_c$ | 0.009 | .0484 | 0.193 |
| Prolate applying bias field along major axis | 0 | $9.3 \times 10^{-4}$ | 0.025 | 0.03 |
| | 0.5 $H_c$ | .0017 | .034 | 0.0465 |
| | 0.75 $H_c$ | .0032 | .05 | .0653 |
| | **0.875 $H_c$** | **0.0063** | **.0662** | **0.09** |
| | $H_c$ | .0018 | .025 | .075 |
| Oblate applying bias field along minor axis | **0.25 $H_c$** | **.94** | **.68** | **1.4** |
| | **0.5 $H_c$** | **0.843** | **0.582** | **1.51** |
| | **0.625 $H_c$** | **0.76** | **0.45** | **1.71** |
| | **0.75 $H_c$** | **0.645** | **0.46** | **1.44** |
| | $H_c$ | .0934 | .0921 | 0.95 |

Finally, we take this best-case nanoparticle (oblate ellipsoid at $H_{dc}=0.625H_c$ with SNR=1.71) and investigate if the SNR can be further improved by applying a narrow band filter around 42.5 MHz. The rationale is that the magnetization response driven by the magnetic field due to proton spin precision at 1 Tesla applied DC field is dominant around 42.5 MHz while the magnetization fluctuations driven by thermal noise is broad band as evidenced by the singe-sided amplitude spectrum shown in Figure 4(a). When a band-pass filter (42-44) MHz was applied, the SNR improved to ~8.

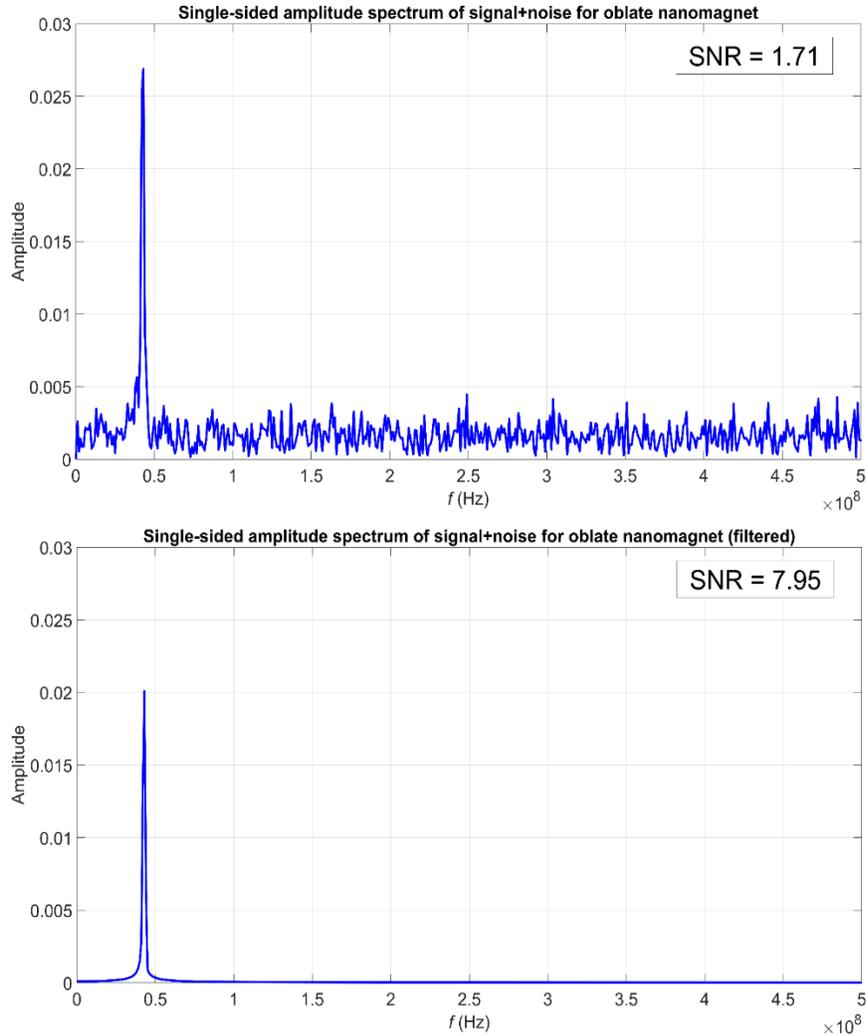

FIG. 4. (a) Frequency spectrum of signal + noise, before filtering (b) Frequency spectrum of signal + noise, after filtering. Both cases for oblate nanomagnet.

**Scaling of SNR with coil core nanoparticle number**

While the SNR with an AC magnetic field of 10 Oe (equivalent of 1mT) shows a SNR ~8 for optimal conditions, this was assuming a single nanomagnet. However, for a large number n of nanomagnets within the core, the magnetization response increases as n (as more nanomagnets coherently contribute to more magnetic moment and therefore greater induced voltage in the coil) according to Equation 3, while the magnetic noise would only increase as $\sqrt{n}$ (as the thermal noise induced fluctuations of nanomagnets have a random phase) according to Equation 5, leading to a SNR increase of $n/\sqrt{n} = \sqrt{n}$.

Now consider, a square detection coil (for ease of calculation) of 2 cm side. Considering the pitch between nanomagnets ~200 nm to avoid significant dipole coupling 10 billion nanomagnets can be accommodated in a single layer of 2 cm by 2 cm. Additionally, as the single nanoparticle layer thickness is ~6 nm the average distance between two such layers can be ~25 nm. Thus, 400,000 such magnetic nanoparticle layers can be accommodated in 1 cm coil thickness. Consequently, $n = 40 \times 10^{14}$ nanomagnets can be incorporated into the sensing coil leading to an increase in SNR from 8 to $\sim 5 \times 10^8$. In other words, with a SNR of 5, one could detect an AC magnetic field of $1/(10^8)$ mT, i.e. an AC magnetic field of 10 pT. Hence, insertion of such nanomagnets could allow detectability to ~10 pT and perhaps even better depending on the density with which nanoparticles that can be inserted in the detection coil. However, it should be noted that pinning sites, inhomogeneities, etc can impede magnetization dynamics [47], create a phase difference, etc, thus decreasing detectability.

The key point is merely filling the detection coil with a soft (high permeability) core would not help as the core would be saturated under the high DC bias fields used for MRI. However, by using appropriate geometry nanoparticles that are still responsive to AC fields from proton precession under strong DC fields, the coil detection sensitivity can be enhanced.

**Conclusion**

Our numerical investigation of nanoparticle magnetization dynamics and LLG of room temperature thermal moment fluctuations confirm the initial zero temperature proposal for nanoparticle based amplification of NMR signal. Such consideration of the thermal fluctuation allows us to predict not just signal amplification values, but realistic room temperature SNR values. Our analysis suggests specific proposal for using oriented anisotropic oblate ellipsoid particles as optimal for achieving SNR improvements over conventional air-filled MRI coils. Much will depend on the quality of the particles used in the coil core: shape uniformity, quality of particles orientation within the core, smoothness of the particles and surface pinning sites (that degrade the effect of magnetization dynamics), and uniformity of the nanoparticle aspect ratio (which determines where the particle has a peak in transverse susceptibility). Further consideration would have to be made of the effect of magnetic particles on the field uniformity within the nuclear spin sample that is being imaged, since such field distortions will broaden the sample spin resonance and will have to be address in both the MRI scanner bore designs that incorporate the nanoparticles within the coils, as well as in the pulse sequences that deal with such inhomogeneous broadening. Nevertheless, our results provide further strong impetus for relatively simple modifications to existing MRI inductive detection coils for achieving improved SNR in scanners operating in 0.1T to 2T polarizing fields range. This promise of a higher SNR would allow for shorter MRI scan time, more compact MRI systems, lower operating fields, and higher accessibility.

# Supplemental Information

# MODELING OF THERMAL MAGNETIC FLUCTUATIONS IN NANOPARTICLE ENHANCED MAGNETIC RESONANCE DETECTION

Tahmid Kaisar[1], Md Mahadi Rajib[1], Hatem ElBidweihy[2], Mladen Barbic[3] and Jayasimha Atulasimha[1*]

[1]*Dept of Mechanical and Nuclear Engineering, Virginia Commonwealth University, Richmond, VA, USA*

[2]*United States Naval Academy, Electrical and Computer Engineering Department, Annapolis, MD, USA*

[3]*NYU Langone Health, Tech4Health Institute, New York, NY 10010, USA*

\* Corresponding author: jatulasimha@vcu.edu

In the main paper, we had plotted the normalized magnetization ($M/M_s$) for (i) a sinusoidal magnetic signal of 800 A/m (10 Oe) amplitude and (ii) thermal noise. However, these plots were provided for only:

1. Two selected cases of DC bias magnetic field for prolate nanomagnet with bias along minor axis; and
2. Two selected cases of DC bias magnetic field for oblate nanomagnet with bias along minor axis.

However, in the supplement we provide figures for normalized magnetization ($M/M_s$) for (i) a sinusoidal magnetic signal of 800 A/m (10 Oe) amplitude and (ii) thermal noise; for all cases of DC bias magnetic fields for:

1. All cases of DC bias magnetic field for prolate nanomagnet with bias along minor axis (Fig S1);
2. All cases of DC bias magnetic field for prolate nanomagnet with bias along major axis (Fig S2); and
3. All cases of DC bias magnetic field for oblate nanomagnet with bias along minor axis (Fig S3).

These supplementary plots form the basis for Table 2 as well as the Signal to Noise Ratio (SNR) plot Fig 3 (e) in the main paper.

## Prolate Nanomagnet with Bias along Minor Axis

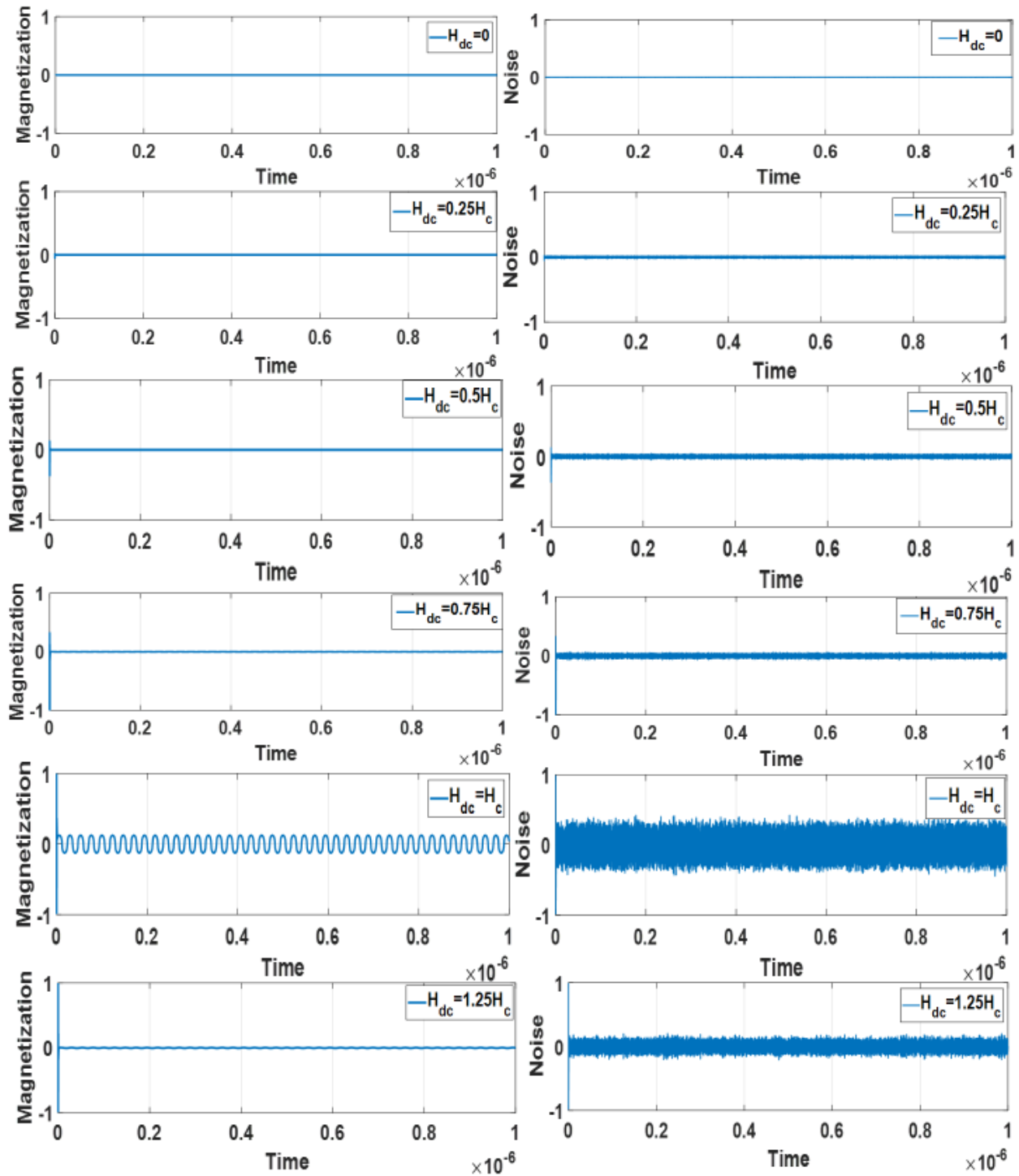

**Fig S1.** Normalized magnetization in a prolate nanomagnet for different bias fields along minor-axis: For sinusoidal magnetic signal of 800 A/m (10 Oe) amplitude without thermal noise (plots on left) and corresponding case with thermal noise only and no signal (plots on right).

## Prolate Nanomagnet with Bias along Major Axis

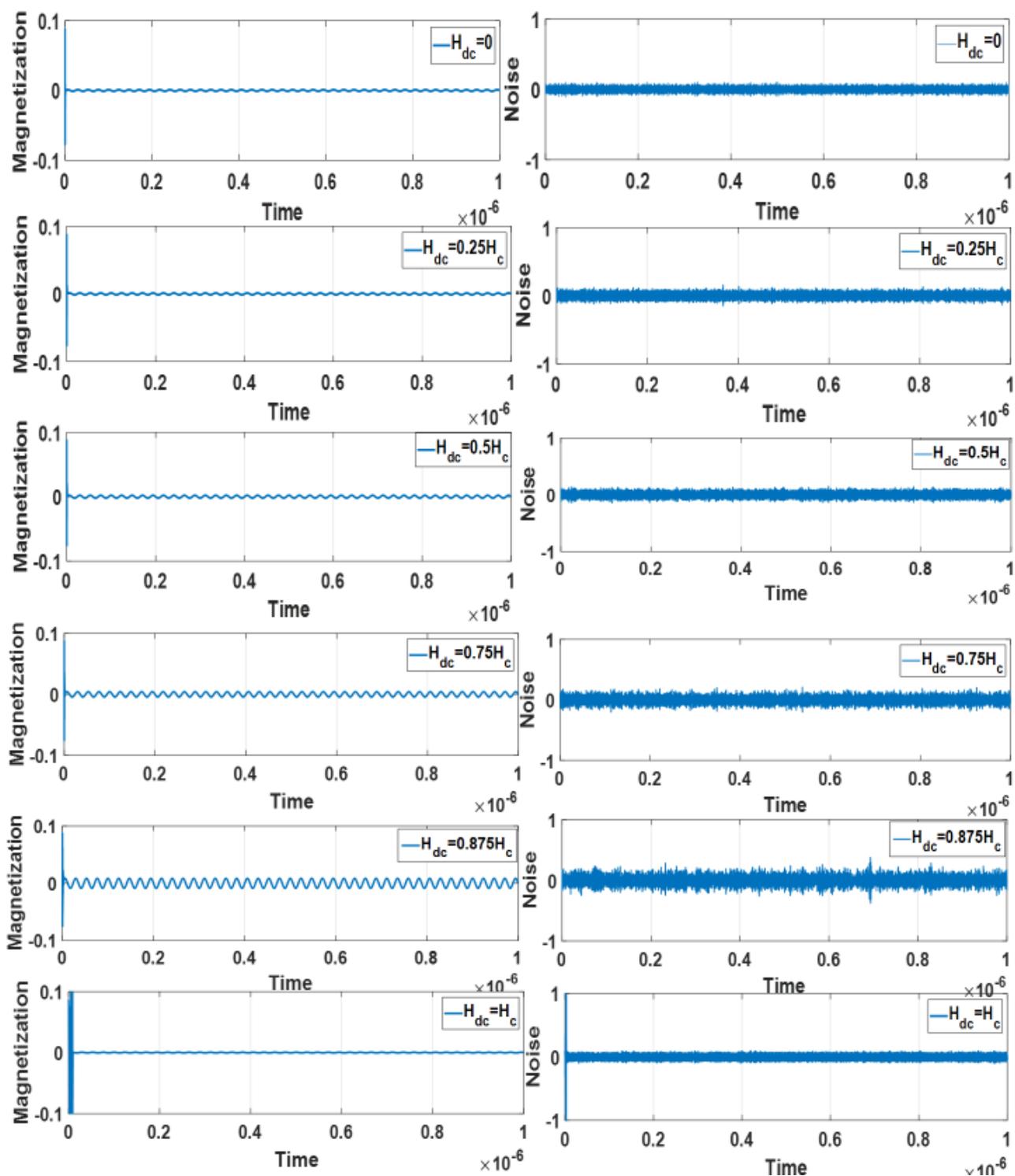

**Fig S2.** Normalized magnetization in a prolate nanomagnet for different bias fields along major-axis: For sinusoidal magnetic signal of 800 A/m (10 Oe) amplitude without thermal noise (plots on left) and corresponding case with thermal noise only and no signal (plots on right).

## Oblate Nanomagnet with Bias along Minor Axis

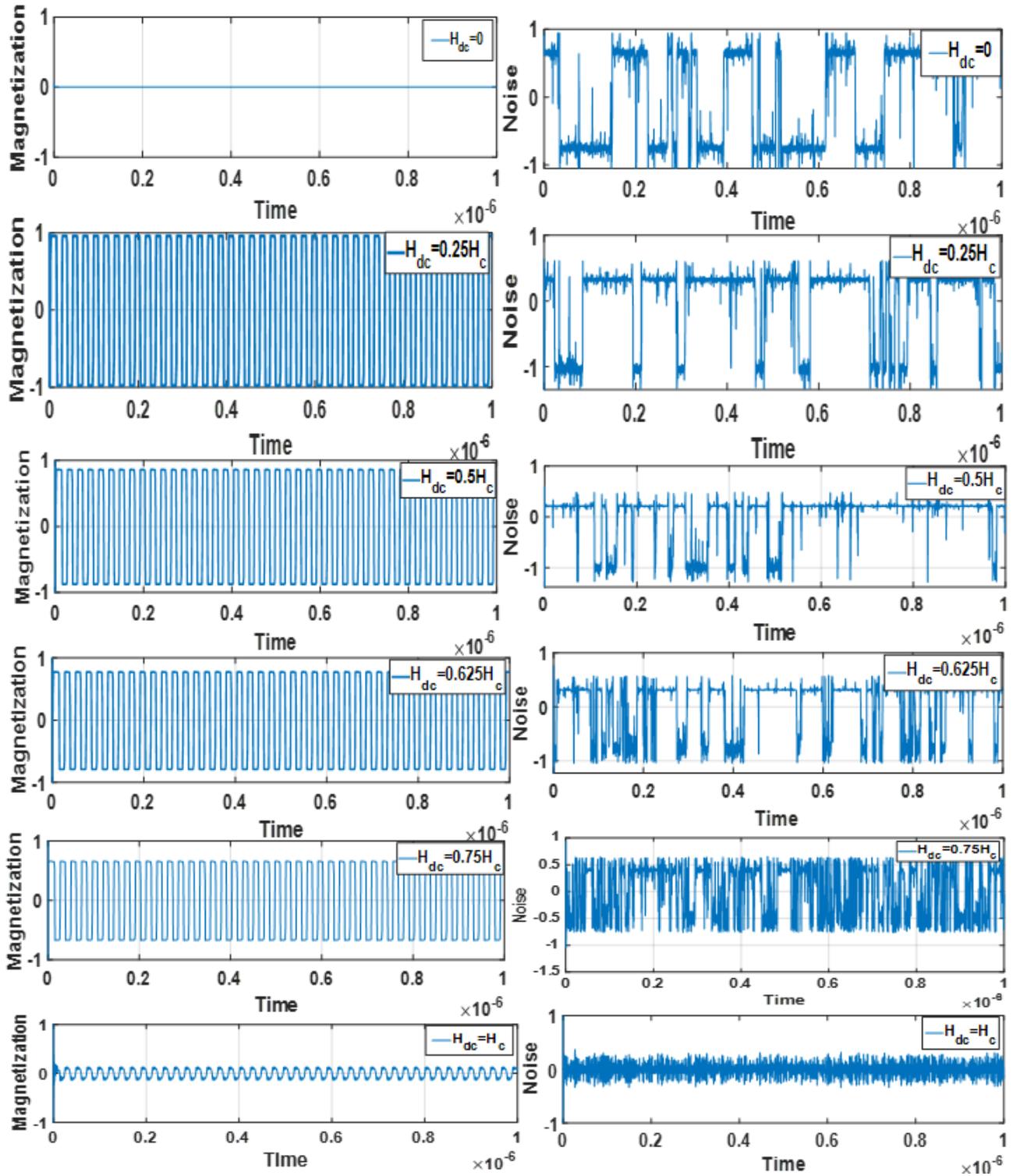

**Fig S3.** Normalized magnetization in an oblate nanomagnet for different bias fields along minor-axis: For sinusoidal magnetic signal of 800 A/m (10 Oe) amplitude without thermal noise (plots on left) and corresponding case with thermal noise only and no signal (plots on right).